# Super-intelligent society for the silver segment: Ethics in design

Jaana Leikas [1], Rebekah Rousi [2], Hannu Vilpponen [3] and Pertti Saariluoma [3]

[1] *VTT Technical Research Centre of Finland, Tampere, Finland*
[2] *University of Vaasa, Vaasa, Finland*
[3] *University of Jyväskylä, Jyväskylä, Finland*

#### Abstract
A super-intelligent AI- society should be based on inclusion, so that all members of society can equally benefit from the possibilities new technologies offer in everyday life. At present, the digital society is overwhelming many people, a large group of whom are older adults, whose quality of life has been undermined in many respects by their difficulties in using digital technology. However, this 'silver segment' should be kept involved as active users of digital services and contribute to the functioning and development of a super-intelligent, AI-enabled society. The paper calls for action-oriented design thinking that considers the challenge to improve the quality of life, with an emphasis on ethical design and ethical impact assessment.

## 1. Introduction

In a techno-optimistic world, digitalization has been seen as the magic wand for the efficiency and accessibility of public services. Society 5.0, introduced by Japan, refers to a super-intelligent society, presented as a solution to building a more sustainable society [1]. It is a response to societal challenges, such as rapid and increasing ageing of society [2,3]. The goal of Society 5.0 is to create a human-centered society in which technological development is truly inclusive. In Society 5.0, physical and virtual spaces are integrated to improve people's quality of life, for example, in health care [1]. They enable the generation of e.g., personal real-time physiological and environmental data, which can be used to actively design people's behavior in society. [2]

Japan is often seen as an example of easing the pressure of ageing demographics. The idea behind Society 5.0 is to support social well-being and a fulfilling life. Both Finland and Japan recognize the importance of technology in easing the burden of demographic ageing. However, Japan stands out in its emphasis on technology and strategic social planning as a whole. The focus is therefore not merely on technological development, but on social development, where a networked society integrates digital tools into people's lives: from what information people need and how they position themselves as customers, to the organization of wellbeing services. To successfully deploy these tools, Society 5.0 will also focus on training people to use digital tools effectively.

Finland has rapidly adopted a 'radical digitalization' approach which has involved the transition of many public services to the digital environment and the disappearance of physical offices. For decades public discourse has referred to a reasoning that due to the loss of active taxpayers, and increase in imminent healthcare costs incurred via ageing, alternative welfare measures must be found [4]. The widespread digitalization together with the emergence of Artificial Intelligence (AI) is expected to bring new ways to streamline services and shape the future, particularly through Internet of Things (IoT) solutions and healthcare applications [5].

However, in addition to forcing users or paying customers to actively participate in the delivery of services – learning and negotiating online systems to ensure a smooth outcome – digital systems are vulnerable to disruption, error, and fraud. They fail in usability and accessibility planning and are vulnerable to cybersecurity risks. They change established ways of acting and operating, and are often developed separately from the context, the real users and understanding of everyday life.

Studies have found that digitalization has polarized society between capable and non-capable and the digital divide experienced by older people has grown [6,7,8,9,10]. In addition, as already seen in recent infamous ethical AI cases [11,12,13,14]. AI algorithms can create social risks and threats. They not only have the power to discriminate and harm (physically, socially and psychologically), but they can, in their application, change social structures and cause significant threat to privacy, security, justice and human autonomy. Data leaks, misuse of data, discriminatory interpretations resulting from inaccurate or skewed data, and excessive direction of service users' attention and actions. It is already apparent that, for example on social media, older people are easily duped by various financial and romantic scams [15].

Drawing on thought behind the design of technological products, and the ethics implicated in particular design and systemic decisions, the current paper focuses on recent developments of digitalization and its impact on the quality of life of older people. We look at the digital societal transformation and ethical issues arising in an evolving technological landscape. In this landscape relationships between humans and technology are rapidly changing from agent-user and object-system to agent-system and increasingly object-users (e.g., via data collection). We observe the tensions between the logics of 'forced digitalization' and attempts to maintain wellbeing among an ageing population.

The ideal of Society 5.0 challenges current design thinking. The transition from traditional technology-intensive design thinking to genuinely holistic design paradigms, where human life and improving its quality are taken as the main goal of design, requires a renewal of design thinking. In this paper, we discuss the nature of the new approach to designing the lives of older citizens.

## 2. Infiltration of digital technology in everyday life

Digitalization has increased in all sectors of society. For example, the health and wellbeing sector has moved towards the digitalization of many services. These include e.g., appointment self-booking, assessing symptoms and service needs, non-urgent communication with professionals, digital questionnaires, coaching, recording measurements (health and fitness data), renewing prescriptions, and reviewing personal data. These emerging digital systems, together with the integration of AI, are anticipated to provide a better customer experience. From a practical and economic perspective, benefits are predicted for both customers and healthcare professionals [16].

In Finland, the digitalization of public services is a national effort [17]. With increasing societal pressures to support older adults in living as independently as possible, many initiatives in remote and digital healthcare have been implemented. For instance, a reform of public social and healthcare services took place in 2023, when the responsibility for organizing services was transferred from the municipalities to 21 established self-governing wellbeing counties [18]. The initiative was an attempt to hinder the pace of the rising expenses faced by the current conditions of an ageing population and economic uncertainty. The use of technology and digitalization of services was seen as an essential part of this development.

Yet, despite these initiatives, the role of technology in the areas of care and wellbeing has much been debated across international scholarly fields including gerontology [19], sociology [20], health sciences [21], psychology [22], and ethics [23,24,25]. Questions pertaining to the role of technology, and particularly in the addressing of social and psychological needs, are becoming ever more imminent. This is not purely based on technological development itself, but alongside the current climate of uncertainty and conditions faced in a world where state financing of health and social services is becoming ever more limited.

As the world turns more unstable, infiltration of digital technology in human life is becoming ever rifer. Trust becomes an immeasurable commodity when the foundations of the understood world (how people experience the world) are balanced upon digitally based information systems [26], increasingly enabled by AI [27]. Trust in decision-makers and application providers is thus a prerequisite for the acceptance of technologies for older people, and for ethically and socially sustainable AI deployment [28]. Trust, however, can erode on numerous dimensions ranging from the ability to use the systems, to the quality and accuracy of information, and intentions behind the technology that is developed and implemented [29].

## 3. The silver segment struggles with digitalization

The silver segment, or silver economy, is a term used to describe the growing population of individuals aged 55 years of age and over, who aspire towards a 'good life' [30]. 'Silver segment' refers to the aging population of 55 years old and above who are active and engaged in society, equipped with wealth, experience, and purchasing power. This market segment, the 'silver economy', is seen as important for businesses as very attractive and promising [31]. However, it is still very underdeveloped in terms of product and service offerings [30].

Although people in their 60s and 70s are generally comfortable with digital devices and applications, it is not always easy for them to take up digital services, particularly the ones required for living and participating in a welfare society. Members of populations aged 55 years old and above were not born and raised using personal computers, Internet, and mobile IT [32]. They remember times of landline telephones, dial up Internet modems, and paper archives. The silver segment belongs thus to a completely different generation of technology users than the generations that will follow them. The technology generation view [33] means that their understanding of how technologies (current and future) are used is built on knowledge of the technologies that were typical of their generation in young adulthood [33]. Typically, this was an era of tangible on-off user interfaces.

The COVID-19 pandemic ironically triggered some positive effects in the rapid digital technology uptake due to individuals of all generations wanting to maintain social contact [34]. Yet, increased interaction with connected digital systems also introduces new problems such as vulnerabilities to cyber-crime, disinformation, and misinformation [35]. Aspects such as multiple applications and platforms, frequently updating software, cyber secure authentication, and security awareness, as well as basic usability of everyday digital systems pose major challenges [36], increasing technological anxiety [37]. Reasons for not using digital services include problems e.g., with vision, physical and cognitive functioning, living in remote areas and lack of support [38]. These challenges cause divides between people as well as between individuals and services [39]. While increasingly more older people utilize digital technology, there is no guarantee of the level of digital literacy (the ability to critically read and evaluate digital content [40], in an era of rapid content advancements - many of which stem from developments in AI.

From a social perspective, disadvantages are not only experienced among the elderly, and this is not simply a 'technology acceptance' challenge. Rather, in addition to providing functional, civic and utilitarian operations, public and societal services have also fulfilled a social role. For instance, while directing individuals towards their digital devices to remain home longer and receive services that were formerly carried out by other human beings seems like the ideal economic solution, the social component of service exchange immediately is lost. What happens here is twofold: 1) the very act of losing this person-to-person contact means increased physical and social isolation among the elderly; and 2) the functional aspects of the person-to-person contact, i.e., contextual awareness, human intelligence, intuition and the ability to flexibly problem-solve on behalf of clients, means increased anxiety and stress when engaging with self-service systems [41,42]. In other words, when there is no one around, who can be asked? Another concrete ethical aspect that arises during this time of inflated prices coupled with radical, self-servitization pertains to where private individuals' money is going.

As described above, Finland is aiming for a situation where half of all basic services can be delivered remotely using digital services by 2025 [43]. Digital technologies can play an assistive, compensatory, and enabling role in the lives of older people, allowing them to live independently and safely at home, thereby reducing public health expenditure. The intense digitalization of core services in society, however, has led to the demise of many traditional face-to-face services and shift to online-only services. At the same time, at least 39% of Finns need frequent help with the internet or digital devices [44]. When technology is difficult to access, some people may be left on the sidelines in a way that does not reflect the principles of a welfare or wellbeing society (where the quality of human life is valued). For those who can use remote services, concerns include the rise of cybercrime and the use of scams to hack into people's banking and identity numbers. Finns lost more than €32 million to digital scammers in 2022 [45]. Practice has shown that people's sensitive health data has even been used as a tool for extortion. Senior citizens are particularly at risk.

For the ageing population, there can be many problems in embracing digitalization, even if an increasing number of individuals are interacting with digital technologies daily. Problems with sensory functions, memory and cognitive functioning, physical capacity, and self-direction can prevent or make it difficult to use digital services [46]. Many people feel that they are unable to manage their affairs through digital services in the same way as during face-to-face encounters.

In addition to everyday services that can be purchased and used independently, the combination of digitalization and AI will also enable the monitoring of older people in their homes by public actors. As the pressure to emphasize home care is increasing, public actors are considering the use of new monitoring technologies and AI applications. Sensor technologies together with AI offer new opportunities for 24/7 health and functioning monitoring at home. Combining AI and health data provides an opportunity to use technology for disease prevention, early detection of symptoms and optimal care of home care clients [47].

An increasingly common scenario is where sensor technology is installed in an elderly person's home, enabling a smart environment to be built that collects information on how well the resident is performing their normal activities. This data, along with many other devices that collect well-being data (such as smart mattresses and well-being bracelets), can be used to detect and visualize fluctuations in the resident's activity level and to see acute or gradual changes in their daily habits. This data complements traditional health information and provides a holistic view of a person's daily performance. Yet, once again from ethical and privacy-related perspectives, we encounter issues that have been raised for decades regarding 'assistive technology' for the ageing people [48]. These include considerations of human dignity, autonomy, and the right to be left alone, as well as questions on control over one's own life [47,49].

## 4. Regulation and informed consent procedure alone are not enough to guarantee the ethical use of AI

The use of registers and electronic health record systems in research, healthcare, and government databases to develop AI services offers countless opportunities to address research and innovation issues related to the well-being and functional ability of older people. Data collection is increasingly needed for e.g., proactive health monitoring, where the aim is to provide a more accurate assessment of a person's functional capacity, behavior as well as communicate information about vital circumstances (i.e., falling, unconsciousness, location etc.). The maintenance of a register is usually required by law and the processing of data by a public authority cannot be based on the consent of the data subject.

The aim is to collect huge amounts of data from multiple data collection sources upon which digital AI-driven systems can adapt, adjust, and tailor output and assistance that is appropriate for specific situations and individuals in question [50] (whether 'user' or 'non-user stakeholder'). The collected mass data (big data) is aggregated and processed to offer a holistic picture of situations and circumstances in question. This is more effective and detailed than utilizing data that has been collected for a single purpose and by an individual application. The more data, the better services can be designed and targeted to individuals. In addition to data collected by public registrars, there will be an increasing need for data collected by individuals themselves on their wellbeing. If, on the other hand, people do not see the collection of data as important, are not able to use the necessary devices or applications, or do not trust the ability of data authorities to treat data confidentially, development will not proceed as expected.

The collection and analysis of mass data in preventive health care brings new perspectives to the concept of informed consent [51]. The resulting data can be used for a variety of purposes, and individuals do not always know how their data is being used, even if they have legally given consent [52]. Among other things such as ensuring understanding and vigilance while reading or receiving the privacy information, the difficulty in giving 'informed' consent is that even the data collector may not know in advance all the purposes for which the data will be used.

As there is confusion regarding the interpretation of the laws on primary and secondary use of health data, the European Health Data Space EHDS [53] in Europe aims to clarify this and contribute to addressing the challenges of accessing and sharing electronic health data. A legislative proposal adopted by the EU Parliament in April 2024 will allow national legislation to provide further elucidation. The benefits of EHDS include empowering individuals to manage their health data, supporting the use of health data to improve healthcare services, research, innovation, and decision-making, and providing the EU with the means to safely harness the potential for the seamless exchange, use and re-use of health data [53]. Finland is a pioneer in the secondary use of social and health data and in 2019 established a data authorization authority for the social and health sector (Findata) to provide guidance, grant permits for secondary use of social and health data and to ensure that the aggregation and sharing of authorized data takes place in a secure manner [54].

Regulatory compliance and informed consent appear to be feasible steps towards designing for privacy and ethical data engagement. However, they are not enough to describe to users and other implicated stakeholders what kinds of data collection is allowed or justified, and why certain types of data are collected. This regulatory approach serves as a seeming safeguard yet does nothing to explain how the data is processed, used (or sold), and how the user or stakeholder may be affected by the collection and use of their data. This is where the paradox comes into play - to provide better, more accurate, more efficient, and relevant services for individuals, the systems need to collect and access more data. However, the design challenge is, how to access and use this data without infringing on an individual's privacy (i.e., gaining access to domains of a person's life of which they do not want to disclose), and, in a way that will protect their identity, autonomy and integrity. This may be challenging in cases of assistive technology in the home or elderly care facilities, particularly if an individual is deemed as incapable of making authoritative decisions over critical areas of their own life (i.e., in cases of memory illness).

In the following, we discuss ethical design of AI-enabled systems for the silver segment.

## 5. Ethical design of AI-enabled systems for the silver segment: basic questions

Digital and AI services for the silver segment is characterized by a systemic whole, where different actors are interconnected within socio-technical entities. Resonating with the sociological Actor Network Theory [55] this web-like effect of contemporary online systems means that everything is dependent on and affects everything else in some way or another. What may seem like a small glitch such as a usability error in the homepage of an internet banking system, may result in people losing trust and finding alternative banks or means of operating their finances. Therefore, when designing digital and AI services consideration must be placed on the complex and dynamic interactions between technology, service providers, older people, and society, how technology affects these actors and how the new AI culture will change human ecosystems.

In the following, we consider two fundamental questions relevant to design, which emphasize an understanding of the term 'good life' - everyday life that is ethically good, well-balanced, fulfilling and in which all basic and higher needs are met in some way [56,57]. These are: i) How to use technology? and 2) How can the use of technology be governed?

### 5.1. How to use technology?

The first approach emphasizes **bridging the digital divide** by investigating socio-technical solutions that have the potential to overcome obstacles related to usability, user experience, and from there privacy and security issues. A valid approach to better embrace the digital divide is **action-oriented design** [58]. Technical artefacts, user interfaces and ubiquitous technologies should all support the achievement of the functional goal as well as possible. In line with the Society 5.0 ideal, action-oriented design helps understand how people can best use particular tools (technologies) at their disposal. Thus, AI as a tool may be instrumental in bridging the usability gap for ageing individuals. Yet, as mentioned earlier its implementation may lead to bigger problems relating to e.g., data vulnerabilities. This is why focus on **how** technology can and should be used is paramount. For, it is not the issue of whether a device or system is available and can be used, it is whether or not it should be used in particular actions.

An essential part of action-oriented design for the silver segment, is user interface design, highlighted by the demand from an ageing population. Age-related changes in physical and cognitive ability mean that digital terminals, typically with small touchscreens and requiring password remembering, are difficult for many older adults to use. As we age, our brains age and our senses deteriorate, which means that it takes longer to learn new things. Thus, achieving an outcome should not cause unnecessary stress for the user and design outcomes should motivate people to use, learn, or recommend them. At best, digital services should improve the quality of people's lives, rendering them essential for fostering a 'good life'. However, as pointed out earlier, the design of accessibility and usability of digital services often falls short. Not all older people have the access, capacity or means to use digital services. This raises issues of discrimination and inequality. In the worst-case scenario, service development will further widen the already existing digital divide, leaving older people excluded from society and services.

In action-oriented design, ethics plays a specific role at different stages of the design process. When defining human actions and their components related to achieving specific goals through the use of digital artefacts, they must be ethically acceptable and aim to improve quality of life. By this we mean, that design accountability must be placed on both what the technology does and is designed for (intention and motivation), as well as what it can be potentially used for - will the technology be prone to use by those with ill-intention to the detriment of the good life? The positive ethical aspects of new action models should thus be tested on several levels. The new practices should genuinely improve the quality of older people's lives by eliminating foreseeable harms or opening new possibilities. It is essential to check that the silver segment can easily use new action models, and that main processes as well as their subprocesses fulfil norms of techno ethics [29]. The action is ethically acceptable only if all the sub-actions are acceptable.

## 5.2. How can the use of technology be governed?

The second question concerns the idea that the development of digital services and AI-powered systems should be based on understanding how people can and wish to live with technology, not only how to use it. Designing digital and AI solutions must consider the life of older people in a holistic way, reflecting on the good life [56].

Anticipating the impact of AI and digitalization in everyday life is not only a task for developers, but also an essential part of the work of policymakers. It is important to look at AI in a wider socio-technical context: through the short and long-term impact of AI on older people's daily lives. Current public policies do not sufficiently prevent the social risks and threats posed by AI [59]. Public administrations should develop the capacity to anticipate and identify AI risks in a holistic way and respond to emerging problems in an agile manner. Responsible public governance guides AI developers and users to ensure that AI delivers for the common good of individuals, communities, and society as a whole. Digitalizing services and increasing utilization of AI challenges public authorities to rethink governance mechanisms so that the need for adoption and implementation of ethics and human rights into policy-making and organizational practices can be fulfilled.

To maximize the positive impact, public institutions should be supported to foster fair and transparent processes to harness AI ethically for the common good. This includes inclusion and empowerment, involving the relevant social actors in an ongoing, open dialogue on desirable and undesirable outcomes, and providing civil society actors with the necessary skills to understand key aspects of ethical use of AI [60,61]. Inclusion in the design process promotes empowerment, which suggests that older people are relevant "co-producers" of life quality [47,62].

The important questions here pertain to the influence on and participation in a socially just distribution of services. Empowerment goes hand-in-hand with "new consumerism", where the silver segment is seen as autonomous consumers with different needs and differences in lifestyles. Here, the individual's right to pursue their own interpretation of happiness and quality of life is a core idea, which in turn, highlights the issues of autonomy and control, as well as issues of fairness and just distribution.

Leikas et al [63] present a framework for designing ethically acceptable public services for older people. The approach is based on three steps complementary to each other: (A) Assessment of change needs in service production; (B) Value mapping of services; and (C) Ethical assessment of service production (Leikas et al. 2020). The first step, assessment of the change needs in service production, is carried out with a cross-section of stakeholders. The idea is to build a shared understanding of future challenges and ethical issues related to acceptance, delivery, and exploitation of emerging technologies in a specific context. Here, relevant stakeholders, their roles and the goals of actions are identified, as well as expectations for future technology.

In the second step, the focus is on understanding a service context and ecosystem from the point of view of fulfilled, lost and new values [64]. This helps actors in embedding responsibility and ethics into the core of the business model through improved understanding of the value proposition.

The third step is to understand what ethical principles and values should define the boundaries of the technology. This deliberation can be carried out with the help of a socio-technical scenario, a discussion tool to capture the different relevant aspects of the service, actors and technology bound to the scenario [65]. Socio-technical scenarios can also be used to broaden stakeholder understanding of different roles in shaping the future, as well as awareness of stakeholder interdependence [66]. With the help of scenarios, it is easier to operationalize 'good' in the design concepts from the point of view of actors, actions and goals of actions, and thus systematically assess the ethical value of the design outcomes [67].

In the public sector context, as can be seen from the example of home monitoring technology, ethical questions are raised by the systemic nature of services. A technology-enabled service scenario for older people may involve both formal and informal care and multiple stakeholders: home care staff, technology company employees, caregivers and service organization representatives. This raises questions of integrity, autonomy and personal privacy of an older person. [63]

## 6. Discussion

The pursuit towards an AI society and the socio-technical changes it will bring challenge technological design thinking to evolve and respond more comprehensively to the needs of an increasingly technological world. When it comes to technology uptake, there is still a lack of knowledge regarding digital wellbeing, its promotion, maintenance, and balance in this unfolding AI-driven digital landscape. Knowledge on potential ethical issues and clear visions are needed to guide the choice of technological ends [68]. This should be based on the core values of the welfare society: inclusion, openness, trust, and equality [69]. Failure to understand and account for the ethical implications of techno-social intervention can result in greater societal costs and ill-being in the long term.

From the perspective of the silver segment, the development of a digital, intelligent society as described in Society 5.0 is not only a solution to the cost and resource problems of an ageing society, but also an important ethical issue of inclusion. This is due to the fact that it can enable older people to participate in society as equal and active members. Society as a whole will benefit: the life experience and tacit knowledge of older people in several areas of life is a resource that we cannot afford not to exploit.

Can the digital divide ever be bridged? It is a mistake to assume that only older people have fallen into the digital divide and young people are digital natives. For many young people, too, using search engines, for example, can be problematic. Although some of the new capabilities of AI may be able to replace and improve previous user interfaces, for example with the help language technology [70] there will be those who, for one reason or another, do not have access and will be left behind. Understanding new technologies such as AI will require effort and learning for everyone. It is the role of society to ensure that citizens of all ages have the opportunities, support, and tools to do so.

If, on the other hand, digital services are difficult to use because of poorly designed user interfaces, inappropriate (unmatching) logic, and shady algorithms because of a lack of concern for people of all ages, including older people with deteriorating physical and cognitive abilities, this is unethical and morally unacceptable. Those who decide on digital services must then focus better on service design and consider the lifestyles of older people in a holistic way.

Coeckelbergh [71] talks about digital humanism, which aptly illustrates the ultimate aspiration of designing for good life. The aim is to develop an intelligent society in harmony with humanistic values, striving for a good life and democracy. As we have shown in our arguments on ethical design, this requires effective dialogue between service developers and public authorities and the inclusion of older people in the debate. A forward-looking, ethical and inclusive approach presented in this paper, ensures that the needs of older adults as well as future society are met in a socially sustainable way.

## 7. References

[1] Fukuyama, M. (2018). Society 5.0: Aiming for a new human-centered society. Japan Spotlight, 27(5), 47-50.
[2] Bartoloni, S., Calò, E., Marinelli, L., Pascucci, F., Dezi, L., Carayannis, E., Revel, G.M., & Gregori, G.L. (2022). Towards designing society 5.0 solutions: The new Quintuple Helix - Design Thinking approach to technology. Technovation, Elsevier, vol. 113(C). https://doi.org/10.1016/j.technovation.2021.102413
[3] UN (2024). United Nations Development Programme Sustainable Development Goals. Sustainable Development Goals | United Nations Development Programme (undp.org)
[4] Watson, A. R. (2016). Impact of the digital age on transforming healthcare. Healthcare WHO World Health Organization (1994). Statement developed by WHO Quality of Life Working Group. In WHO Health Promoting Glossary 1998. WHO/HPR//HEP/98.1.
[5] Greco, L., Percannella, G., Ritrovato, P., Tortorella, F., & Vento, M. (2020). Trends in IoT based solutions for health care: Moving AI to the edge. Pattern recognition letters, 135, 346-353.
[6] Arieli, R., Faulhaber, M.E., & Bishop, A.J. (2023). Bridging the Digital Divide: Smart Aging in Place and the Future of Gerontechnology. In: Ferdous, F., Roberts, E. (eds) (Re)designing the Continuum of Care for Older Adults. Springer, Cham. https://doi.org/10.1007/978-3-031-20970-3_1
[7] Blažič, B.J., & Blažič, A.J. (2020). Overcoming the digital divide with a modern approach to learning digital skills for the elderly adults. Educ Inf Technol 25, 259–27. https://doi.org/10.1007/s10639-019-09961-9
[8] Choudrie, J., Pheeraphuttranghkoon, S., & Davari, S. (2020). The Digital Divide and Older Adult Population Adoption, Use and Diffusion of Mobile Phones: A Quantitative Study. Inf Syst Front 22, 673–695. https://doi.org/10.1007/s10796-018-9875-2
[9] Francis, J., Ball, C., Kadylak, T., & Cotten, S.R. (2019). Aging in the Digital Age: Conceptualizing Technology Adoption and Digital Inequalities. In: Neves, B.,Vetere, F. (eds) Ageing and Digital Technology. Springer. https://doi.org/10.1007/978-981-13-3693-5_3
[10] Frydman, J.L., Gelfman, L.P., Goldstein, N.E., et al. (2022). The Digital Divide: Do Older Adults with Serious Illness Access Telemedicine? J GEN INTERN MED 37, 984–986. https://doi.org/10.1007/s11606-021-06629-4
[11] Bao, Y., Li, W., Ye, Y., & Zhang, Q. (2022). Ethical Disputes of AI Surveillance: Case Study of Amazon. In 2022 7th International Conference on Financial Innovation and Economic Development (ICFIED 2022) (pp. 1339-1343). Atlantis Press.
[12] Elish, M. C. (2019). Moral crumple zones: Cautionary tales in human-robot interaction (pre-print). Engaging Science, Technology, and Society (pre-print).
[13] Feffer, M., Martelaro, N., & Heidari, H. (2023). The AI Incident Database as an Educational Tool to Raise Awareness of AI Harms: A Classroom Exploration of Efficacy, Limitations, & Future Improvements. In Proceedings of the 3rd ACM Conference on Equity and Access in Algorithms, Mechanisms, and Optimization (pp. 1-11).
[14] Nasim, S. F., Ali, M. R., & Kulsoom, U. (2022). Artificial intelligence incidents & ethics a narrative review. International Journal of Technology, Innovation and Management (IJTIM), 2(2), 52-64.
[15] Wnek-Gozdek, J. (2020). Prevention connected with romantic relationships in the Internet. International Journal of Innovation, Creativity and Change. Volume 14, Issue 11, 2020.

[16] Ma, B., Yang, J., Wong, F. K. Y., Wong, A. K. C., Ma, T., Meng, J., ... & Lu, Q. (2023). Artificial intelligence in elderly healthcare: A scoping review. Ageing Research Reviews, 83, 101808.
[17] Finnish Government (2024). Finland's National Roadmap: EU Digital Decade Policy Programme 2030. Publications of the Finnish Government 2024:7. http://urn.fi/URN:ISBN:978-952-383-743-0
[18] STM (2023). The Ministry of Social Affairs and Health. Wellbeing services counties will be responsible for organising health, social and rescue services. https://stm.fi/en/wellbeing-services-counties
[19] Broekens, J., Heerink, M., & Rosendal, H. (2009). Assistive social robots in elderly care: a review. Gerontechnology, 8(2), 94-103.
[20] Pedersen, I., Reid, S., & Aspevig, K. (2018). Developing social robots for aging populations: A literature review of recent academic sources. Sociology Compass, 12(6), e12585.
[21] Sun, X., Yan, W., Zhou, H., Wang, Z., Zhang, X., Huang, S., & Li, L. (2020). Internet use and need for digital health technology among the elderly: a cross-sectional survey in China. BMC public health, 20, 1-8.
[22] Rogers, W. A., & Fisk, A. D. (2010). Toward a psychological science of advanced technology design for older adults. Journals of Gerontology Series B: Psychological Sciences and Social Sciences, 65(6), 645-653.
[23] Abdel-Keream, M. (2023). Ethical challenges of assistive robotics in the elderly care: Review and reflection. Robots in care and everyday life, 121.
[24] Johnston, C. (2022). Ethical design and use of robotic care of the elderly. Journal of Bioethical Inquiry, 19(1), 11-14.
[25] Sharkey, A., & Sharkey, N. (2012). Granny and the robots: ethical issues in robot care for the elderly. Ethics and information technology, 14, 27-40.
[26] Oehler, A., & Wendt, S. (2018). Trust and financial services: The impact of increasing digitalisation and the financial crisis. In The return of trust? Institutions and the public after the Icelandic financial crisis (pp. 195-211). Emerald Publishing Limited.
[27] Etienne, H. (2021). The future of online trust (and why Deepfake is advancing it). AI and Ethics, 1(4), 553-562.
[28] Sutrop, M. (2019). Should we trust artificial intelligence? A Journal of the Humanities and Social Sciences, 23(4), 499–522.
[29] Saariluoma, P., & Rousi, R. (2020). Emotions and technoethics. In R. Rousi, J. Leikas & P. Saariluoma (Eds.), Emotions in Technology Design: From Experience to Ethics (pp. 167-189). Cham: Springer
[30] Kohlbacher, F., & Herstatt, C. (eds.). (2011). The silver market phenomenon: Marketing and innovation in the aging society. Second Edition. Springer.
[31] Lipp, B., & Peine, A. (2022). Ageing as a driver of progressive politics? What the European Silver Economy teaches us about the co-constitution of ageing and innovation. Ageing and Society, 1–13. doi:10.1017/S0144686X22000903
[32] Bennett, S. (2012). Digital natives. In Encyclopedia of cyber behavior (pp. 212-219). IGI Global.
[33] Docampo Rama, M. (2001). Technology Generations – Handling Complex User Interfaces. Eindhoven: University of Eindhoven.
[34] Sin, F., Berger, S., Kim, I. J., & Yoon, D. (2021). Digital social interaction in older adults during the COVID-19 pandemic. Proceedings of the ACM on Human-Computer Interaction, 5(CSCW2), 1-20.
[35] Nash, S. (2019). Older adults and technology: moving beyond the stereotypes. https://longevity.stanford.edu/older-adults-and-technology-moving-beyond-the-stereotypes/
[36] Rajamäki, J. (2021). Ethics of Cybersecurity in Digital Healthcare and Well-Being of Elderly at Home. In Proceeding of the 20th European Conference on Cyber Warfare and Security ECCWS 2021. Academic Conferences International.
[37] Frishammar, J., Essén, A., Bergström, F., & Ekman, T. (2023). Digital health platforms for the elderly? Key adoption and usage barriers and ways to address them. Technological Forecasting and Social Change, 189, 122319.
[38] Saukkonen P, Kainiemi, E, Virtanen, L, Kaihlanen A-M, Koskinen S, Sainio P, Koponen P, Kehusmaa S, Heponiemi T. (2022) Non-use of Digital Services Among Older Adults During the

Second Wave of COVID-19 Pandemic in Finland: Population-Based Survey Study. In: Gao, Q., Zhou, J. (eds) Human Aspects of IT for the Aged Population. Design, Interaction and Technology Acceptance. HCII 2022. Lecture Notes in Computer Science, vol 13330. Springer.

[39] Mubarak, F., & Suomi, R. (2022). Elderly forgotten? Digital exclusion in the information age and the rising grey digital divide. INQUIRY: The Journal of Health Care Organization, Provision, and Financing, 59, 00469580221096272.

[40] Pangrazio, L., Godhe, A. L., & Ledesma, A. G. L. (2020). What is digital literacy? A comparative review of publications across three language contexts. E-learning and Digital Media, 17(6), 442-459.

[41] Lian, J. W. (2021). Why is self-service technology (SST) unpopular? Extending the IS success model. Library Hi Tech, 39(4), 1154-1173.

[42] Shiwen, L., Kwon, J., & Ahn, J. (2022). Self-service technology in the hospitality and tourism settings: A critical review of the literature. Journal of Hospitality & Tourism Research, 46(6), 1220-1236.

[43] Yrittäjät (2018). Suomi 2025. Digitalisaatio ja digipolitiikka 2025: Lähdekoodi uudelle ja uudistuvalle yrittäjyydelle. (In Finnish). https://www.yrittajat.fi/wp-content/uploads/2021/07/sy_digitalisaatio_ja_digipolitiikka_2025-1.pdf

[44] DVV (2022). Digital skills Report 2022. Digital and population data services agency. https://dvv.fi › Digitaitoraportti_2022_ENG

[45] DVV (2023). Watch out, verify, warn others: Noticeable increase in digital scams. Press Release, Digital and population data services agency. https://dvv.fi/en/-/watch-out-verify-warn-others-noticeable-increase-in-digital-scams-in-july-december-2022?languageId=en_US

[46] Heponiemi, T., Kainiemi, E., Virtanen, L., Saukkonen, P., Sainio, P., Koponen, P., & Koskinen, S. (2023). Predicting Internet Use and Digital Competence Among Older Adults Using Performance Tests of Visual, Physical, and Cognitive Functioning: Longitudinal Population-Based Study. Journal of Medical Internet Research, 25, e42287.

[47] Leikas, J., & Kulju, M. (2018). Ethical consideration of home monitoring technology: A qualitative focus group study. Gerontechnology 2018;17(1):38-47; https://doi.org/10.4017/gt.2018.17.1.004.00.

[48] Yusif, S., Soar, J., & Hafeez-Baig, A. (2016). Older people, assistive technologies, and the barriers to adoption: A systematic review. International journal of medical informatics, 94, 112-116.

[49] Ienca, M., & Fosch-Villaronga, E. (2019). Privacy and security issues in assistive technologies for dementia. Intelligent Assistive Technologies for Dementia: Clinical, Ethical, Social, and Regulatory Implications, 221.

[50] Boyd, D., & Crawford, K. (2012) Critical questions for big data. Information, Communication & Society, 15:5, 662-679. http://doi.org/10.1080/1369118X.2012.678878

[51] Froomkin, A.M. (2019). Big Data: Destroyer of Informed Consent. Yake Journal of Health Policy, Law, and Ethics, Preprint, University of Miami Legal Studies Research Paper Forthcoming. https://papers.ssrn.com/sol3/papers.cfm?abstract_id=3405482

[52] Mittelstadt, B.D., & Floridi, L. (2016). The Ethics of Big Data: Current and Foreseeable Issues in Biomedical Contexts. Sci Eng Ethics 2016; 22, 303–341. https://doi.org/10.1007/s11948-015-9652-2

[53] Council of Europe. (2024). Proposal for a Regulation on the European Health Data Space - Analysis of the final compromise text with a view to agreement. https://www.consilium.europa.eu/media/70909/st07553-en24.pdf

[54] Findata (2024). Social and Health Data Permit Authority. https://findata.fi/en/

[55] Latour, B. (2007). Reassembling the social: An introduction to actor-network-theory. Oup Oxford.

[56] Saariluoma, P., Cañas, J.J., & Leikas, J. (2016). Designing for Life - A human perspective on technology development. Palgrave MacMillan. ISBN 978-1-137-53046-2 DOI 10.1057/978-1-137-53047-9

[57] Leikas, J. (2009). Life-Based Design - A holistic approach to designing human-technology interaction. VTT Publications 726. Edita Prima Oy. http://www.vtt.fi/inf/pdf/publications/2009/P726.pdf. ISBN 978-951-38-7374-5

[58] Saariluoma, P., Parkkola, H., Honkaranta, A., Leppänen, M., & Lamminen, J. (2009). User psychology in interaction design: the role of design ontologies. Future interaction design II, 69-86.

[59] Jobin, A., Ienca, M., & Vayena, E. (2019). The global landscape of AI ethics guidelines. Nature Machine Intelligence, 1(9), 389-399.
[60] Sigfrids, A., Nieminen, M., Leikas, J., & Pikkuaho, P. (2022). How should public administrations foster the ethical development and use of artificial intelligence? A review of proposals for developing governance of AI. Front. Hum. Dyn. 4:858108. https://doi.org/10.3389/fhumd.2022.858108
[61] Winfield, A. F., & Jirotka, M. (2018). Ethical governance is essential to building trust in robotics and artificial intelligence systems. Philos. Trans. Royal Soc. 376, 20180085. /10.1098/rsta.2018.0085
[62] Sanders, E.B.N., & Stappers, P.J. (2008). Co-creation and the new landscapes of design. CoDesign 2008;4(1):5-18. https://doi.org/10.1080/15710880701875068
[63] Leikas, J., Sigfrids, A., Stenvall, J., & Nieminen, M. (2020). Good Life Ecosystems – Ethics and Responsibility in the Silver Market. In: Rauterberg M. (Ed.) Culture and Computing. HCII 2020. Lecture Notes in Computer Science, vol 12215. Springer, Cham, pp. 105-122. https://doi.org/10.1007/978-3-030-50267-6_9
[64] Palomäki, K., & Rana, P. Mapping multiple stakeholder value in service innovation: an industrial case study. International Journal of Services Sciences, 6(3/4), 218 (2017).
[65] Carroll, J.M. (2000). Five reasons for scenario-based design. Interact. Comput. 13, 43–60.
[66] Lucivero, F. Ethical Assessments of Emerging Technologies: Appraising the Moral Plausibility of Technological Visions; The International Library of Ethics, Law and Technology, Volume 15, Springer, Heidelberg (2016).
[67] Leikas, J., Koivisto, R., & Gotcheva, N. Ethical framework for designing autonomous systems. J. Open Innov. Technol. Mark. Complex. 5, 18 (2019), doi:10.3390/joitmc5010018.
[68] Rousi, R., Vakkuri, V., Daubaris, P., Linkola, S., Samani, H., Mäkitalo, N., ... & Abrahamsson, P. (2022, August). Beyond 100 ethical concerns in the development of robot-to-robot cooperation. In 2022 IEEE International Conferences on Internet of Things (iThings) and IEEE Green Computing & Communications (GreenCom) and IEEE Cyber, Physical & Social Computing (CPSCom) and IEEE Smart Data (SmartData) and IEEE Congress on Cybermatics (Cybermatics) (pp. 420-426). IEEE.
[69] Tsamados, A., Aggarwal, N., Cowls, J., Morley, J., Roberts, H., Taddeo, M., & Floridi, L. (2022). The ethics of algorithms: key problems and solutions. AI & SOCIETY, 37(1), 215-230.
[70] Sayers, D., Sousa-Silva, R., Höhn, S., Ahmedi, L., Allkivi-Metsoja, K., Anastasiou, D., ... & Yayilgan, S. Y. (2021). The Dawn of the Human-Machine Era: A forecast of new and emerging language technologies.
[71] Coeckelbergh, M. (2024). Why AI Undermines Democracy and What To Do About It. Cambridge: Polity.